\documentclass[aps,pra,twocolumn,10pt,shortbibliography]{revtex4-1}
\usepackage{amsmath}
\usepackage{amsfonts}
\usepackage{amssymb}{\tiny }
\usepackage{mathtools}
\usepackage{times}
\usepackage{bm}
\usepackage{cases}
\usepackage[version=4]{mhchem}
\usepackage{graphicx}
\usepackage[caption=false]{subfig}
\usepackage[usenames]{color}
\usepackage[bookmarks=true,colorlinks,linkcolor=blue, urlcolor=blue,citecolor=blue]{hyperref}

\definecolor{darkred}{rgb}{0.847059, 0.141176, 0.164706}
\definecolor{darkgreen}{rgb}{0,0.4,0}
\definecolor{darkblue}{rgb}{0.254902, 0.411765, 0.882353}

\allowdisplaybreaks%
\newcommand{\bs}{\boldsymbol}
\newcommand{\mc}{\mathcal}

\newcommand{\dg}{\dagger}
\newcommand{\pg}{{\phantom{\dagger}}}

\begin{document}
\title{Some experimental schemes to identify quantum spin liquids}
\author{Yong Hao Gao$^{1}$}
\author{Gang Chen$^{2,1}$}
\affiliation{$^{1}$State Key Laboratory of Surface Physics and Department of Physics, Fudan University, Shanghai 200433, China}
\affiliation{$^{2}$Department of Physics and HKU-UCAS Joint Institute for Theoretical
and Computational Physics at Hong Kong, The University of Hong Kong, Hong Kong, China}

\date{\today}

\begin{abstract}
Despite the apparent ubiquity and variety of quantum spin liquids in theory, 
experimental confirmation of spin liquids remains to be a huge challenge. 
Motivated by the recent surge of evidences for spin liquids in a series of 
candidate materials, we highlight the experimental schemes, involving the 
thermal transport and spectrum measurements, that can result in smoking-gun 
signatures of spin liquids beyond the usual ones. 
For clarity, we investigate the square lattice spin liquids and theoretically 
predict the 
possible phenomena that may emerge in the corresponding spin liquids candidates. 
The mechanisms for these signatures can be traced back to either the intrinsic 
characters of spin liquids or the external field-driven behaviors. 
Our conclusion does not depend on the geometry of lattices and can 
broadly apply to other relevant spin liquids.	
\end{abstract}
\maketitle

\section{Introduction}
\label{sec1}

The search for exotic states in quantum magnets has been a central topic of intensive investigation in modern condensed matter physics. Among the various novel quantum states, quantum spin liquid (QSL), a non-symmetry-breaking phase beyond conventional Landau paradigm, is particularly appealing  due to its potentially relevant to high-temperature superconductivity~\cite{Anderson1987} and quantum-computation  applications~\cite{KITAEV20062}, in which the localized spins are highly entangled and remain disordered even down to zero temperature~\cite{Balents2010,RevModPhys.89.025003,Savary2016}. The concept of QSL was originally proposed by Anderson in 1973 when he studied the nearest neighbor antiferromagnetical Heisenberg model on triangular lattice  ~\cite{Anderson1973}.  Although the true ground state of this model has turned out to be a $120^\circ$ magnetically ordered state, it does ignite the investigations of QSLs and the interplay between frustration and quantum fluctuation. Theoretically, various QSL ground states has been proposed, which are usually characterized by fractional spinons strongly coupled to emergent gauge field. In particular, Kitaev proposed~\cite{KITAEV20062} an exactly solvable spin-$1/2$ model on the honeycomb lattice in 2006, in which the presence of bond-dependent Kitaev interactions induces strong quantum fluctuations frustrating the spin configurations and resulting in a Kitaev QSL state.

From the experimental point of view, the kagom\'e, hyperkagom\'e 
and pyrochlore lattice materials with corner-sharing geometries 
or the edge-sharing  triangular lattice  materials provide ideal 
platforms to realize such an exotic magnetic ground state. In 
most QSL candidates, the results from measurements such as 
magnetization, heat capacity and nuclear magnetic resonance are 
consistent with properties of QSLs~\cite{Balents2010,RevModPhys.89.025003,Savary2016} 
and show no onset of long-range order at low temperatures. 
Besides, the crucial signature of a QSL is the presence of 
deconfined and fractionalized spinons that can be directly 
measured by inelastic neutron scattering and revealed in 
the excitation continuum, which is fundamentally different 
from the sharp and coherent magnon modes in ordered magnets. 
The magnetic excitation continuum indeed has been observed 
in geometrically frustrated spin-$1/2$ systems with both 
two-dimensional (2D) and three-dimensional (3D) lattices~\cite{Han2012,Shen2016,Balz2016,Gao2019}. 
However, it has been shown that a simple spectral continuum 
may also originate from a spin glass state or disorder-induced 
state~\cite{PhysRevLett.120.087201, PhysRevX.8.031028,PhysRevLett.119.157201}. 
Thus most experimental evidences so far are not strong enough 
to completely confirm a QSL.

\begin{figure}[b]
	\centering
	\includegraphics[width=8.1cm]{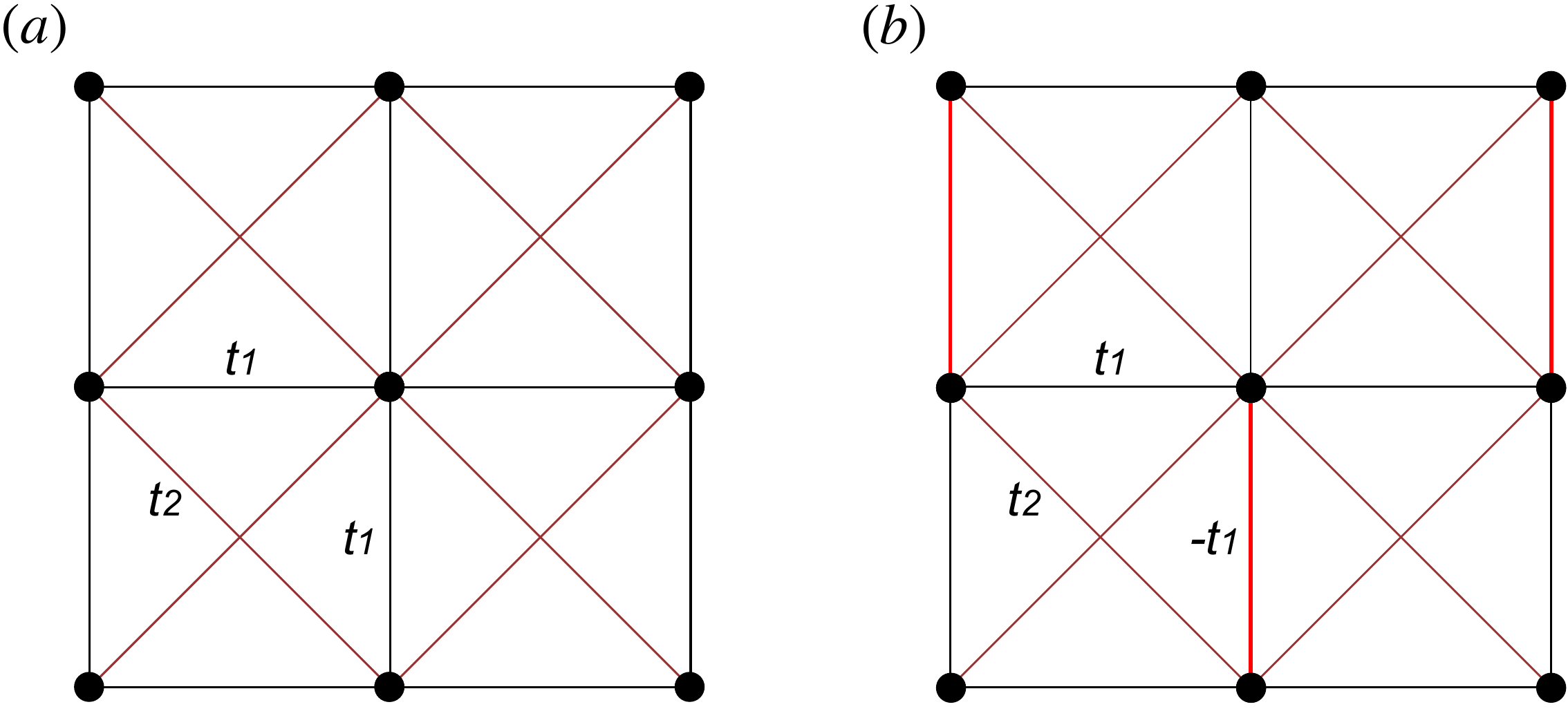}
	\caption{Schematic illustration of spinon hoppings up to second neighbors on the square lattice. 
	(a) The zero-flux QSL with a uniform nearest-neighbor spinon hopping coefficient ${t_{1,ij}=t_{1,ji}=t_1}$ 
	and 	next-nearest-neighbor spinon hopping coefficient ${t_{2,ij}=t_{2,ji}=t_2}$. 
	(b) The $\pi$-flux QSL with a gauge fixing such that the red thick lines stand 
	for negative spinon hopping coefficient ${t_{1,ij}=t_{1,ji}=-t_1}$, while the 
	meaning of other lines remains unchanged.}
	\label{fig1}
\end{figure}

Here we highlight the experimental schemes that would give 
smoking-gun signatures of QSLs beyond the usual ones mentioned above, 
including the thermal transport and spectrum measurements. First, 
we note that the $\pi$-flux QSL states would result in an enhanced 
spectral periodicity of the spinon continuum. It is the translation 
symmetry that is intrinsically fractionalized and renders such an
enhanced spectral periodicity~\cite{PhysRevB.65.165113,PhysRevB.87.104406,PhysRevB.90.121102}, 
much analogous to the fractional charge excitation in the fractional 
quantum Hall states where the global U(1) charge conservation gives 
the fractional charge quantum number to the fractionalized 
excitation~\cite{PhysRevB.96.085136}. Second, the Zeeman coupling 
will enter the spinon Hamiltonian under moderate magnetic fields, 
which can lead to an X-shaped crossing of spectrum compatible 
with the splitting spinon bands. The third case is a chiral 
spin liquid (CSL) that would exhibit a quantized thermal Hall effect 
and a gapped spectrum. In this work, we explicitly demonstrate these 
strong and nontrivial experimental signatures by considering square 
lattice QSLs, but we stress that our conclusion does not depend on 
the geometry of lattice and can generally apply to other relevant QSLs.

The remaining parts of the paper are structured as follows. 
In Sec.~\ref{sec2}, we introduce the Abrikosov fermion construction 
for the square lattice $J_1$-$J_2$ Heisenberg model and illustrate 
the enhanced spectral periodicity for $\pi$ flux QSL at the mean field level. 
In Sec.~\ref{sec3}, we explain the spectrum crossing under magnetic fields, 
which is well compatible with Zeeman-splitted spinon bands. 
In Sec.~\ref{sec4}, we consider the ring-exchange term in the weak 
Mott insulating regime and show the integer quantized thermal Hall 
effect if the ground state is driven into a CSL phase. We conclude 
in Sec.~\ref{sec5} with a discussion of the results.

\section{Enhanced spectral periodicity of the spinon continuum}
\label{sec2}

The discovery of fractional quantum Hall effect experimentally realized the  
theoretical concepts of emergence and fractionalization~\cite{Tsui1982,Laughlin1983}. 
The QSL is another obvious case of fractionalization~\cite{PhysRevB.65.165113}, 
especially the $\pi$-flux QSL that owns fractionalized translation symmetry, 
and such a  fractionalization would result in an observable phenomenon in experiment. 
For clarity, we begin with a $J_1$-$J_2$ Heisenberg model on the 2D square lattice, 
which has attracted enormous research interests due to its intimate relation 
to the magnetism in high-temperature superconducting materials~\cite{Anderson1987,RevModPhys.78.17}, 
but we would not only constrain ourselves in this model. 
The Hamiltonian of $J_1$-$J_2$ model is given by,
\begin{equation}
\label{Eq: J1J2model}
H = 
    J_1\sum_{\langle ij \rangle} \bs{S}_i\cdot\bs{S}_j 
  + J_2\sum_{\langle\langle ij \rangle\rangle} \bs{S}_i\cdot\bs{S}_j,
\end{equation}
where $ \bs{S}_i$ is the spin-$1/2$ operator at the site $i$, 
${J_1 > 0}$ and ${J_2 > 0}$ are the nearest-neighbor (NN) and 
next-nearest-neighbor (NNN) couplings. Moreover, the sums 
${\langle ij \rangle}$ and ${\langle\langle ij \rangle\rangle}$ 
run over NN and NNN pairs, respectively. Although there is no 
geometrical frustration on a square lattice, by switching on 
an antiferromagnetic $J_2$ term indeed brings competing 
interactions and is expected, with the aid of quantum 
fluctuations, to destroy the conventional antiferromagnetic 
N\'eel state and result in a quantum disordered QSL. In fact, 
for the small $J_2$ region, just as the NN Heisenberg model 
on a square lattice, the ground state is generally believed 
to be a $(\pi,\pi)$ long-ranged N\'eel order. On the other hand, 
when $J_2$ becomes comparable to $J_1$, the $(\pi,0)$ and$(0,\pi)$ 
stripe long-range order is stabilized~\cite{PhysRevB.86.075111,PhysRevLett.87.097201,PhysRevB.98.134410,PhysRevB.88.060402,PhysRevB.86.024424}. 
For the intermediate region ${0.4 \lesssim J_2/J_1 \lesssim 0.6}$, 
it has been interpreted as the magnetically disordered QSL phase 
with either gapless or gapped excitations in various numerical 
studies~\cite{PhysRevB.86.075111,PhysRevLett.87.097201,
PhysRevB.98.134410,PhysRevB.88.060402,PhysRevB.86.024424}. 
We mainly focus on the intermediate disordered regime 
in this work and assume it realizes a QSL.

\subsection{Abrikosov fermion construction}

To analyze the QSL phase of this model, we here adopt the well-known and 
widely-used Abrikosov fermion construction since it can be utilized to 
study both gapped and gapless phases, while the Schwinger 
boson formalism has the limitation to study gapped phases~\cite{PhysRevB.45.12377}. 
In the Abrikosov fermion representation, the effective 
spin-$1/2$ operator $\bs{S}_{i}$ on site $i$ is given by 
\begin{equation}
\label{Eq: AbrRep}
\bs{S}_{i}^{} =
\frac{1}{2} \sum_{\alpha,\beta} f_{i\alpha}^\dagger \bs{\sigma}_{\alpha\beta}^\pg f_{i\beta}^\pg
\end{equation}
where $f_{i,\alpha}^{\dagger} (f_{i,\alpha})$ creates (annihilates) 
a spinon with the spin index $\alpha=\uparrow,\downarrow$ at site $i$, 
and $\bs{\sigma}$ is a vector of three Pauli matrices. The Hilbert
space constraint
${\sum_\alpha f^\dg_{i\alpha} f^\pg_{i\alpha} = 1}$ on local
fermion number is imposed to project out the unphysical states 
and faithfully reproduce the physical Hilbert space. 
Substituting Eq.~\eqref{Eq: AbrRep} into the $J_1$-$J_2$ 
Hamiltonian Eq.~\eqref{Eq: J1J2model}, one obtains an 
interacting four-fermion system. Performing a mean-field 
decoupling would reformulate the interacting fermionic system 
to a quadratic level~\cite{PhysRevB.65.165113}. Specifically, 
by ignoring the pairing channel, the general quadratic spinon 
Hamiltonian with only spinon hopping sector is obtained as
\begin{equation}\label{Eq: Hmf}
\begin{split}
H_{MF}=&-\sum_{i  j,\alpha}(t_{1,ij}f_{i,\alpha}^{\dagger}f_{j,\alpha}+t_{2,ij}f_{i,\alpha}^{\dagger}f_{j,\alpha}+h.c.)\\
&-\mu\sum_{i,\alpha}f_{i,\alpha}^{\dagger}f_{i,\alpha}
\end{split}
\end{equation}
where we have maintained the SU(2) spin rotation symmetry of the original spin model and the local occupation constraint is relaxed such that only its average value satisfies ${\sum_\alpha \langle f^\dg_{i\alpha} f^\pg_{i\alpha} \rangle = 1}$. The global chemical potential $\mu$ is introduced as a Lagrange multiplier to enforce such a constraint. Moreover, the mean-field parameters $t_{1,ij}$ and $t_{2,ij}$ represent the hopping amplitudes between NN and NNN sites, respectively. In the numerical studies, such as variational Monte Carlo approach, a similar mean-field Hamiltonian to Eq.~\eqref{Eq: Hmf} can also be exploited as a good starting point
to construct the many-body variational wavefunction.

Here we stress again our purpose in the following is not to solve 
for the detailed ground state of a spin Hamiltonian as in Eq.~\eqref{Eq: J1J2model}. 
Instead, we assume that the system stabilizes a magnetically disordered 
QSL phase in the intermediate region ${0.4 \lesssim J_2/J_1 \lesssim 0.6}$, 
as suggested by a variety of numerical studies~\cite{PhysRevB.86.075111,PhysRevLett.87.097201,PhysRevB.98.134410,PhysRevB.88.060402,PhysRevB.86.024424}. 
Comparing with pursuing solving a spin model exactly, 
it might be a better, or at least as a supplementary strategy 
to start from the potential QSL states and then single out 
the nontrivial and robust experimental signatures that allow 
us to distinguish a QSL. We then start from a mean-field theory 
to proceed with our analysis since much can already be 
learned from a mean-field investigation. 
A full treatment of the original spin model requires the 
involvement of all quantum fluctuations of the parameters 
around the mean-field solution, but the robust and intrinsic 
experimental signatures could maintain even the fluctuations 
are included.

\begin{figure*}[t]
	\centering
	\includegraphics[width=15cm]{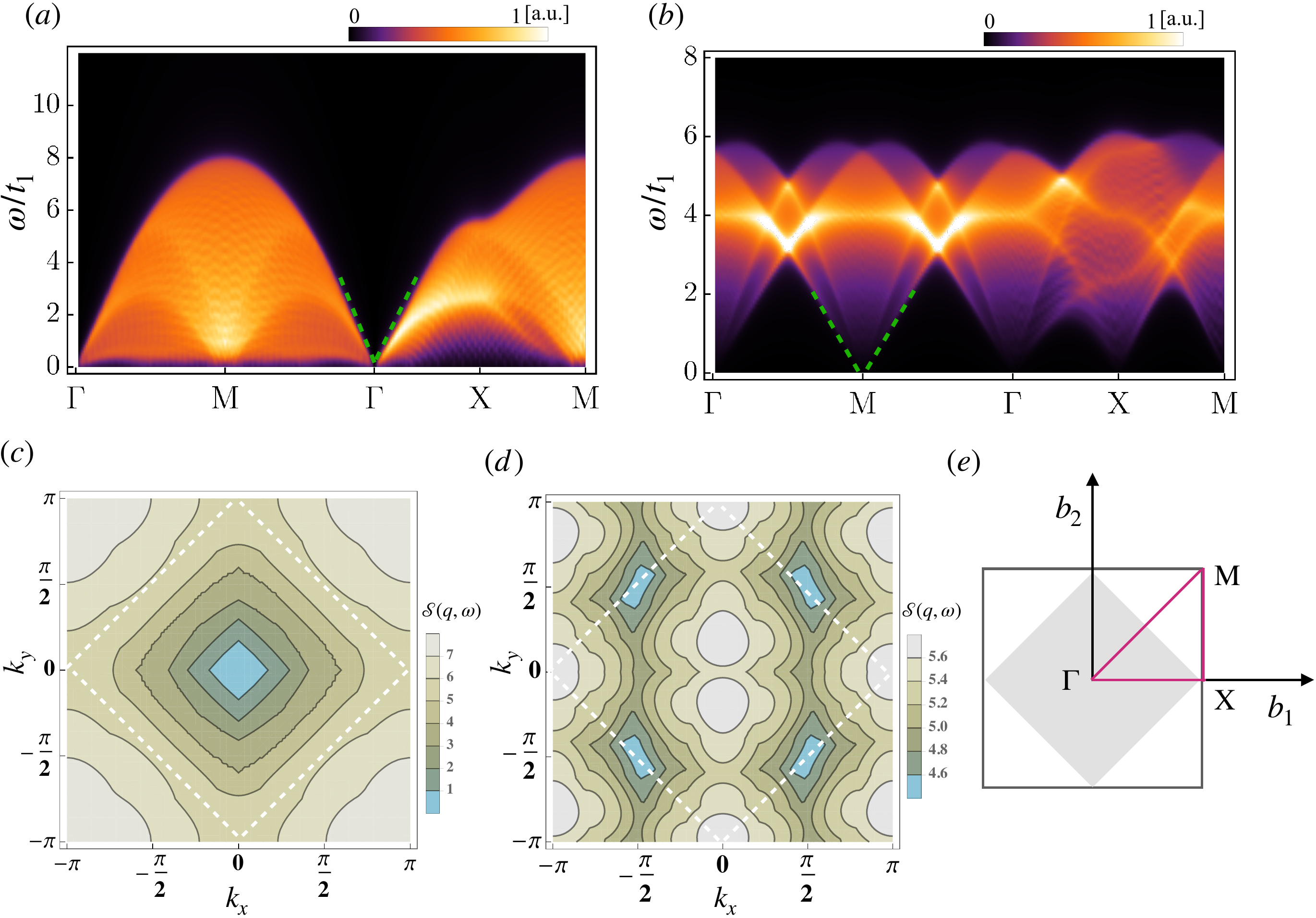}
	\caption{(Color online.) Calculated dynamical spin structure factor 
	$\mc{S}(\boldsymbol{q},\omega)$ 
	along the high symmetry line $\Gamma$-M-$\Gamma$-X-M in the first Brillouin zone, 
	(a) zero-flux spinon Fermi surafce QSL with ``V"-shape character around 
	the $\Gamma$ and (b) $\pi$-flux Dirac QSL with clear low-energy cone features 
	around the high symmetry points. Contour plot of the upper edge of 
	$\mc{S}(\boldsymbol{q},\omega)$ in the first Brillouin zone for 
	(c) Zero-flux spinon Fermi surafce QSL and (d) $\pi$-flux Dirac QSL. 
	(e) Original Brillouin zone (outer black square) and the folded Brillouin zone 
	(light gray square) of square lattice. The parameters adopted in the calculation 
	are ${t_2/t_1=0.2}$ with zero temperature ${k_BT/t_1=0}$.}
	\label{fig2}
\end{figure*}

The spinons fulfill the projective symmetries of the square lattice, 
and the mean-field parameters $t_{1,ij}$ and $t_{2,ij}$, also called 
mean-field ansatzs in the literatures, should be constrained by a 
systematic projective symmetry group (PSG) analysis~\cite{PhysRevB.65.165113}, 
which results in a classification of all possible QSLs. 
It is shown~\cite{PhysRevB.65.165113} that under open boundary condition of 
the square lattice, one can always fix the gauge such that the gauge
transformations for translations satisfy
\begin{equation}
\label{gaugeT}
\mc{G}_{T_1}(i)=1, \quad \mc{G}_{T_2}(i)=\eta^{i_x+i_y}_{xy},
\end{equation}
where $T_1,~T_2$ are two translations of square lattice, and  
$\mc{G}_{T}(i)$ is the associated gauge transformation. The 
spinons thus fulfill the gauge enriched~\cite{PhysRevB.65.165113} 
translations $\tilde{T}_1=\mc{G}_{T_1}T_1$ and $\tilde{T}_2=\mc{G}_{T_2}T_2$. 
The parameter $\eta_{xy}$ takes values $\pm1$, then it divides 
all the possible QSLs on the lattice into two categories, 
i.e., zero-flux states correspond to $\eta_{xy}=+1$ and 
$\pi$-flux states correspond to $\eta_{xy}=-1$. In particular, 
the $\pi$-flux states harbor a fractionalized translation symmetry, 
which would result in a sharp signature in experiment. 
To illustrate this idea, we adopt the simplest two cases 
and the values of $t_{ij}$ in Eq.~\eqref{Eq: Hmf} we have 
taken are shown in Fig.~\ref{fig1}, where Fig.~\ref{fig1} (a) 
corresponds to a zero-flux QSL, while Fig.~\ref{fig1} (b) 
corresponds to a $\pi$-flux QSL. It can be easily verified 
that ${t_{1,ij}=t_{1,ji}=-t_1}$ on the (thick) red bonds in Fig.~\ref{fig1} (b)  
involve a background $\pi$ gauge flux through the lattice, 
while this flux has no influence to the NNN hoppings. 
In fact, Figs.~\ref{fig1} (a) and (b) correspond to a 
spinon Fermi surface QSL and a $\pi$-flux Dirac QSL, respectively.

\subsection{Inelastic neutron scattering spectrum and enhanced spectral periodicity}

Inelastic neutron scattering (INS) measurement represents the best 
experimental probe to \emph{directly} detect the magnetic excitations, 
and the dynamical information of excitations is encoded into the 
dynamical spin structure factor
\begin{equation}\label{Eq: Sqw1}
\begin{split}
\mathcal{S}(\bs{q},\omega)
=&\frac{1}{N}\sum_{i,j}e^{i\bs{q}\cdot(\bs{r}_i-\bs{r}_j)}
\int dt e^{-i\omega t}\langle\bs{S}_i^{-}(t)\cdot\bs{S}_{j}^{+}(0)\rangle\\
=&\sum_{n}\delta[\omega-\xi_n(\bs{q})]|\langle n|\bs{S}^+_{\bs{q}}|G\rangle|^2, 
\end{split}
\end{equation}
where $N$ is total number of lattice sites and the summation 
runs over all the excited eigenstates $|n\rangle$ with 
$\xi_n(\boldsymbol{q})$ being the energy of the $n$-th 
excited state with the momentum $\boldsymbol{q}$,  
while $|G\rangle$ stands for the spinon ground state 
with spinons filling the Fermi sea. In the numerical calculations,  
the delta function is taken to have a Lorentz broadening, 
$\delta({\omega})=\eta/[\pi(\omega^2+\eta^2)]$ with ${\eta=0.1t_1}$. 
Moreover, since ${\bs{S}^+_{\bs{q}}=\sum_{\bs{k}}f_{\bs{k}+\bs{q},\uparrow}^{\dagger}f_{\bs{k},\downarrow}}$, 
the summation in Eq.~\eqref{Eq: Sqw1} should be over all possible 
spin-1 excited states that are characterized by one spinon 
particle-hole pair crossing the spinon Fermi energy with 
a total energy $\omega$ and total momentum $\boldsymbol{q}$~\cite{Shen2016}. 
In other words, the momentum-transfer $\bs{q}$ and energy-transfer 
$\omega(\bs{q})$ of the neutron should be shared between the 
spinon particle-hole pair
\begin{eqnarray}
\bs{q} &=& \bs{k}_1 - \bs{k}_2, \\
\xi(\bs{q}) &=& \omega_1(\bs{k}_1) - \omega_2(\bs{k}_2),
\end{eqnarray}
where $\omega_1(\bs{k})$ [$\omega_2(\bs{k})$] is the spinon (hole) 
excitation energy with momentum $\bs{k}$ and the minus sign comes 
from the nature of hole excitation. The above equations indicate 
the two-spinon spectrum continuum, which is often count as a 
manifestation of fractionalized excitations, is a general character 
of QSLs. In Fig.~\ref{fig2} (a) and (b), we present the density plots 
of dynamical structure factor $\mc{S}(\bs{q},\omega)$ along the high 
symmetry line $\Gamma$-M-$\Gamma$-X-M marked in Fig.~\ref{fig2} (e) 
for zero-flux QSL and $\pi$-flux QSL, respectively. As depicted in 
Fig.~\ref{fig2} (a), besides the obvious spectral continuum, a 
clear V-shape appears around $\Gamma$ point, which originates from 
the particle-hole pair excitations crossing near the Fermi surface~\cite{Shen2016}. 
While for the $\pi$-flux Dirac QSL in Fig.~\ref{fig2} (b), 
the obvious phenomenon becomes the low-energy cone features 
that originate from the inter- (large $\bs{q}$) and 
intra-Dirac (small $\bs{q}$) cone scatterings, which 
is just the reflection of a Dirac QSL with Dirac band touching, 
\emph{not} the $\pi$-flux for the spinons of the QSL. 
However, these signatures and evidences seem not to be 
strong enough to confirm a QSL, as it has been shown 
that a simple spectrum continuum, including the V-shape feature, 
in the dynamical spin structure factor can also be explained 
in the scenarios of usual glassy and  
disorder-induced states~\cite{PhysRevLett.120.087201, PhysRevX.8.031028,PhysRevLett.119.157201}. 
Therefore, additional signature of spectrum that is more unique to QSLs 
is expected to diagnose the fractionalized excitations.

For this purpose, next we consider the intrinsic fractionalized translation 
symmetry due to the $\pi$ flux of the $\pi$-flux QSL. According to Eq.~\eqref{gaugeT}, 
the translation symmetries of spinons should satisfy
\begin{eqnarray}
\tilde{T}_1 \tilde{T}_2 \tilde{T}^{-1}_1 \tilde{T}^{-1}_2 = \eta_{xy},
\end{eqnarray}
where $\tilde{T}_1$ and $\tilde{T}_2$ are gauge enriched translations 
acting on the spinon degrees of freedom instead on the physical spins. 
For zero-flux state with ${\eta_{xy}=1}$, the translations commute as usual, 
while for the $\pi$-flux state with ${\eta_{xy}=-1}$, the anti-commutation 
of translations implies a fractionalization of symmetry. It was first 
realized that the crystal momentum fractionalization of spinons has 
dramatic effects on the neutron spectrum~\cite{PhysRevB.87.104406, PhysRevB.90.121102}. 
As a consequence, the periodicity of the upper excitation edge of 
the dynamic spin structure factor defined by
\begin{eqnarray}
{\rm edge}(\bs q) = \max_{\bs{k}} [\omega_1 (\bs{k}+\bs{q}) -\omega_2 (\bs{k}) ]
\end{eqnarray}
is doubled~\cite{PhysRevB.99.205119}. Therefore, for the zero-flux state, 
the upper two-spinon excitation edge should have the usual periodicity, 
while for the $\pi$-flux state, the upper two-spinon excitation edge 
exhibits an enhanced periodicity, that could serve as a sharp 
identification of fractionalized excitations beyond the simple spectrum 
continuum, since it is impossibly mimicked by any disorder-induced states. 
We illustrate the contour plots of the upper edge of the dynamic 
spin structure for the zero-flux QSL and $\pi$-flux QSL in 
Figs.~\ref{fig2} (c) and (d), respectively. It is clear that 
the $\pi$-flux QSL exhibits a fractionalization 
pattern with an enhanced periodicity in Fig.~\ref{fig2} (d), 
which is readily accessible to the INS measurements. 
The enhanced spectral periodicity with a folded Brillouin zone 
is the dynamical property rather than the static property and 
can not be captured by the static spin structure factor.

\section{Spectrum crossing under magnetic fields}
\label{sec3}

\begin{figure}[t]
	\centering
	\includegraphics[width=8.5cm]{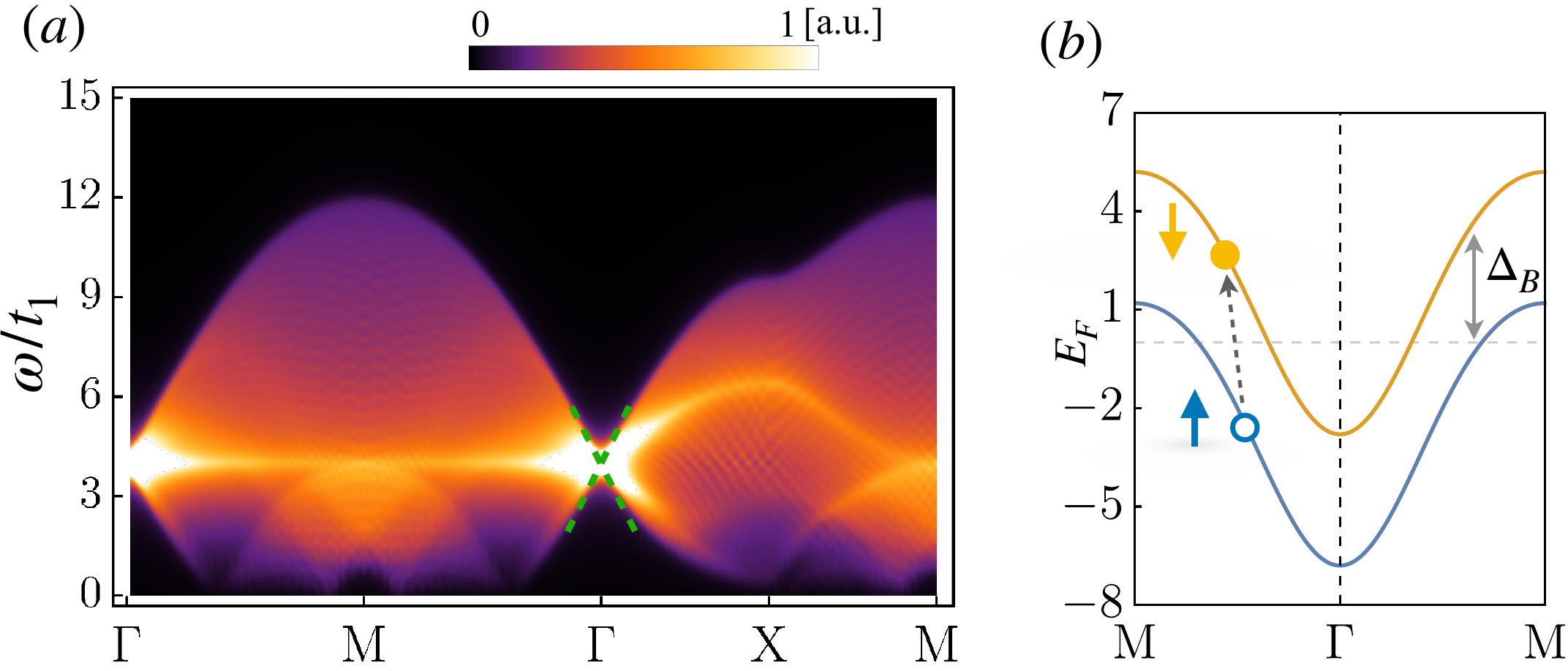}
	\caption{(a) Dynamic spin structure factor for zero-flux QSL with ${t_2/t_1=0.2}$ 
	and $z$-direction magnetic fields ${B_z/t_1= 4}$. 
	(b) Schematic illustration of the particle-hole excitations with small momenta. 
	Such excitations for each $q$ are degenerate at zero field, while the 
	two-fold degeneracy is lifted soon when the Zeeman field is turned on.}
	\label{fig3}
\end{figure}

\begin{figure*}[t]
	\centering
	\includegraphics[width=15cm]{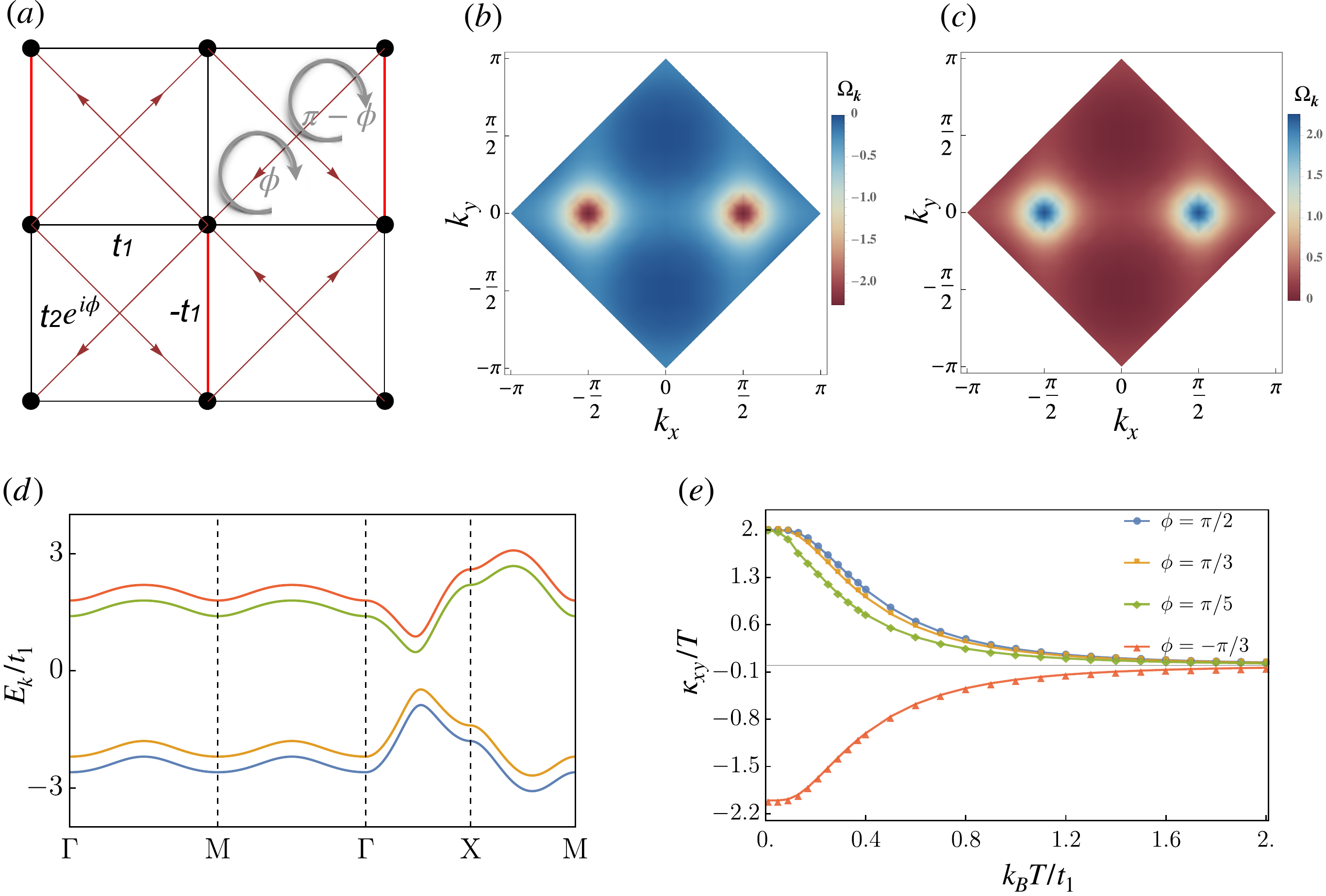}
	\caption{(a) Schematic illustration of the spinon hopping matrix involving 
	the complex second neighbor hopping coefficients, hopping along the 
	arrows corresponds to $\phi$, while hopping oppositely the arrows corresponds 
	to $-\phi$. Contour plot of Berry curvatures calculated when 
	${t_2/t_1=0.3}$ and ${\phi=\pi/2}$ for (b) the lower two bands and 
	(c) the upper two bands. 
	(d) Representive spinon bands calculated when ${t_2/t_1=0.2}$, 
	${\phi=\pi/3}$ and ${B_z/t_1= 0.4}$, 
	the corresponding Chern numbers from the lowest band to the 
	highest one are ${-1, -1, +1, +1}$, respectively. 
	(e) The evolution of thermal Hall conductivity with temperature 
	for different phase $\phi$, where 
	the magnetic field is fixed at ${B_z/t_1= 0.4}$, 
	and the unit of $\kappa_{xy}/T$ here is ${\pi k_B^2/6\hbar}$.
	}
	\label{fig4}
\end{figure*}

In Sec.~\ref{sec2}, we have explicitly demonstrated 
that the excitation spectrum of the $\pi$-flux QSL 
would show an enhanced spectral periodicity in the reciprocal space, 
which is an intrinsic character of this kind of state and would be a 
strong evidence for QSL. Then a natural question is that for the 
zero-flux state, do we have any key features besides the simple 
spectral continuum to distinguish it from the glassy or disorder-induced states? 
In this section, we consider the field-driven behavior of a QSL. 
Under a moderate external magnetic field, the QSL phase should 
not be destroyed immediately, then we can safely consider the 
QSL behavior with fields. The fermionic spinon, unlike the usual 
electron, is electrical charge neutral  
and does not directly couple to the external magnetic field 
through the conventional orbital coupling. 
Especially in the strong Mott insulating regime that we focus in this section, 
there is only~\cite{PhysRevB.96.075105} a simple linear Zeeman coupling
\begin{equation} 
\label{Eq: Zeeman}
H_B=-\frac{B_z}{2}\sum_{i,\alpha \beta} f_{i,\alpha}^{\dagger}
\sigma^z_{\alpha \beta}f_{i,\beta},
\end{equation}
where we have considered the $z$-direction field for concreteness, 
and the Land\'e $g$ factor and Bohr magneton $\mu_B$ have been 
absorbed in $B_z$. When the Zeeman term Eq.~\eqref{Eq: Zeeman} 
enters the spinon Hamiltonian Eq.~\eqref{Eq: Hmf}, a direct 
consequence would be the splitting spinon bands for spin-$\uparrow$ 
and spin-$\downarrow$ spinons, as show in Fig.~\ref{fig3} (b). 
At the zero momentum transfer, there should be a large density 
of spinon particle-hole excitations with the energy ${\omega(\bs{q}=0)=B_z}$ 
due to the splitting bands.

To further observe how these splitting spinon bands reflect 
in the INS spectrum, we utilize the full zero-flux spinon 
Hamiltonian with Zeeman term to calculate the dynamical 
spin structure factor, the result is depicted in Fig.~\ref{fig3} (a). 
We note that the spectrum preserves the spinon continuum and 
remains gapless as expected, while the spectral weight at the 
$\Gamma$ point is strongly enhanced at the band splitting energy 
${\omega=B_z}$, well compatible with the large density of spinon 
particle-hole excitations with the energy $B_z$ at zero momentum transfer. 
Besides the enhancement of the spectral intensity at the $\Gamma$ 
point and the Zeeman splitting energy, the most interesting 
phenomenon is that the V-shaped spectrum around $\Gamma$ point 
under zero field is recast into a X-shaped spectral crossing 
near the Zeeman splitting energy, which is a rather unique 
field-driven behavior for spinons originating from splitting 
spinon bands~\cite{PhysRevB.96.075105}. Therefore, we conclude 
that the X-shaped spectral crossing under Zeeman fields is 
another strong evidence for the QSL with spinon Fermi surface, 
since it is hard to imagine 
that the spin glassy or disorder-induced freezing states could reproduce 
such a spectral crossing under the magnetic fields.

\section{Quantized thermal Hall effect}
\label{sec4}

In the previous two sections, we mainly focused on 
the INS spectral properties of the square lattice QSLs and 
suggest two strong signatures for QSLs, either from the intrinsic 
characters of QSLs or from the external field-driven behaviors. 
This is partly motivated by the fact that the spectral continuum 
of spin excitations detected by INS measurement is a direct 
evidence and consequence for the fractionalization. 
In this section, we turn to the thermal transport properties 
that can unveil the nature of the low-energy itinerant excitations. 
In the QSL phase, the deconfined spinons carry energy and thus 
transport heat under the temperature gradient field, same as the 
electrons transport charge in an electrical field. At the mean-field 
level where the spinons are nearly free, the spinon term should 
dominate the thermal conductivity $\kappa_{xx}$ if spinons 
exist at low energies, and a finite residual 
of $\kappa_{xx}/T$ in the low-temperature limit is proposed 
as the evidence for spinon Fermi surface or gapless spinon excitations, 
since the phonon contribution should be small at low temperatures. 
However, one should note that the total thermal conductivity 
in a Mott insulator (especially the the strong spin-orbit coupled 
Mott insulator) usually is not a simple addition of the magnetic 
contribution and the phonon contribution~\cite{PhysRevB.96.054445}, 
while the mutual scattering between the magnetic excitations 
and the phonons could suppress the value of $\kappa_{xx}$ 
observed in the transport experiments. To obtain the smoking-gun 
signatures of QSLs, here we only focused on the thermal 
Hall effect of spinons as phonons usually do not contribute 
to thermal Hall transport.

Thermal Hall effect in QSLs is a rather nontrivial phenomenon 
since the spinon does not directly couple to external fields 
through conventional Lorentz coupling as we mentioned in Sec~\ref{sec3}. 
In a former work~\cite{Gao2020}, we have pointed out that for 
the non-centrosymmetric U(1) QSLs with Dzyaloshinskii-Moriya (DM) 
interaction, the synergism of a moderate external magnetic field 
and DM interaction could effectively generate an internal U(1) 
gauge flux for the spinons and twists the spinon motion, 
which would result in a spinon thermal Hall effect under 
the temperature gradient field. This mechanism also has 
its limitation, as for the square lattice QSLs and other 
centrosymmetric QSLs, the DM interaction is usually 
prohibited by lattice symmetry~\cite{DZYALOSHINSKY1958241,PhysRev.120.91}. 
An alternative way is to consider the square lattice QSL 
in the weak Mott insulating regime, in which the strong charge 
fluctuations can bring the ring exchange spin interaction, 
and induce a scalar spin chirality term under 
fields~\cite{PhysRevB.51.1922,PhysRevB.73.155115,PhysRevLett.104.066403} 
 \begin{equation}
 \label{hamChi}
H_{\chi}=J_{\chi}\sum_{i,j,k\in\triangle}
\sin{\Phi} \, \bs{S}_i\cdot\bs{S}_j\times\bs{S}_k,
\end{equation}
where the triangle $\triangle$ for sums is formed by three 
neighbor sites involving two NN bonds and one NNN bond as 
shown in Fig.~\ref{fig4} (a), and $\Phi$ is the magnetic flux 
through the triangular plaquette in a counter clockwise way. 
A finite value of this term explicitly breaks the time reversal 
symmetry and parity, while their combination is well preserved. 
Moreover, decoupling this term to the quadratic level would 
induce a complex second neighbor hopping coefficient 
$t_2'e^{i\theta_{ij}}$, thus the total NNN hopping amplitude 
should be
 \begin{equation}
 \begin{split}
t_2+t_2'e^{i\theta_{ij}}
&=(t_2+t_2'\cos{\theta_{ij}})+i t_2'\sin{\theta_{ij}}\\
&=\sqrt{t_2^2+t_2'^2+2t_2t'_2\cos(\theta_{ij})}e^{\phi_{ij}}\\
&=t_2^*e^{\phi_{ij}}
 \end{split}
\end{equation}
where $t_2^*$ is renormalized as $t_2^*=\sqrt{t_2^2+t_2'^2+2t_2t'_2\cos(\theta_{ij})}$, 
and $\phi_{ij}$ is defined by 
$\tan(\phi_{ij})=t_2'\sin{\theta_{ij}}/(t_2+t_2'\cos{\theta_{ij}})$, 
in our case they are both tuning parameters and we can also 
denote the total NNN hopping coefficient as $t_2e^{\phi_{ij}}$ 
for simplicity of notation. A convenient convention of these 
hopping amplitudes is shown in Fig.~\ref{fig4} (a),  
where we have chosen the induced complex NNN hopping 
amplitudes on top of a $\pi$-flux QSL.

In this sense, the significance of external magnetic field $B_z$ is {\sl twofold}. 
It not only provides a linear Zeeman coupling to split the spinon bands, 
but also induces a complex NNN hopping coefficient that breaks the time 
reversal symmetry. Using the hopping matrix marked in  Fig.~\ref{fig4} (a), 
we plot a typical spinon dispersion in Fig.~\ref{fig4} (d), where the fermion 
number constraint guarantees the bands to be half-filled, thus the lowest 
two spinon bands are fully occupied while the upper two spinon bands are 
completely empty, corresponding to a gapped QSL state. 
According to Polyakov's argument for 2D compact U(1) gauge 
theory~\cite{POLYAKOV1977429}, if the state is trivially gapped,  
the dynamical U(1) gauge field will be confined due to the proliferation 
of monopoles and the system should enter a confining ordered state. 
However, the spinon Hamiltonian we considered is indeed nontrivial, 
since these spinon bands own non-vanishing Berry curvatures
 [see Figs.~\ref{fig4} (b) and (c) for the contour plots of 
 the spinon Berry curvatures] and the total Chern number of 
 the lowest two occupied bands is ${C=-2\neq 0}$. Therefore, 
 there would be a Chern-Simons term in the theory for gauge 
 fluctuations and this state can safely get rid of the confinement 
 issue, resulting in a CSL. 
 Theoretically, the chiral edge modes of CSL would
 contribute to an integer quantized thermal Hall effect 
 under temperature gradient field, which is the 
 smoking-gun signature of a CSL.

To explicitly demonstrate the quantized thermal Hall effect 
and its evolution when varying temperature and the NNN hopping 
phase $\phi$, we numerically calculate the thermal Hall 
conductivity for this QSL state. The thermal Hall conductivity 
formula is obtaioned~\cite{PhysRevLett.107.236601} as
\begin{equation}
\label{thermcon}
\kappa_{xy}=-\frac{k_B^2}{T}\int d\epsilon(\epsilon-\mu)^2
\frac{\partial f(\epsilon,\mu,T)}{\partial \epsilon}\sigma_{xy}(\epsilon)\,,
\end{equation}	
where ${f(\epsilon,\mu,T)=1/[e^{\beta(\epsilon-\mu)}+1]}$ 
is the Fermi-Dirac distribution with chemical potential $\mu$, 
and 
\begin{equation}
\label{eq14}
{\sigma_{xy}(\epsilon) 
	= - \frac{1}{\hbar}  
	\sum_{\boldsymbol{k},\xi_{n,\boldsymbol{k}}<\epsilon}\Omega_{n,\boldsymbol{k}}}
\end{equation}
is the zero temperature Hall coefficient for a system 
with the chemical potential $\epsilon$ and Berry curvature 
${\Omega_{n\boldsymbol{k}}=-2 {\rm Im}\langle  
{\partial u_{n\boldsymbol{k}}}/{\partial k_x}|
{\partial u_{n\boldsymbol{k}}}/{\partial k_y}
\rangle}$ for the spinon band indexed by $n$. 
Since in our case the chemical potential $\mu$ lies in the gap, 
in the zero temperature limit, Eq.~\eqref{thermcon}  
is recast into 
\begin{equation}
\frac{\kappa_{xy}}{T}=-\frac{\pi k_B^2}{6\hbar}\sum_{n=1,2}C_n,
\end{equation}
where $C_n$ with $n=1,2$ is the Chern number of the two filled spinon bands. 
In Fig.~\ref{fig4} (e), we plot the evolution of $\kappa_{xy}/T$ with 
temperature at various NNN hopping phase $\phi$, the absolute value 
of them are all monotonically decreasing with increasing temperatures, 
and $\kappa_{xy}/T$ gets smaller for smaller $\phi$ at the same nonzero temperature. 
In particular, the quantized number 2 in the zero temperature limit 
is consistent with theoretical analysis, and its sign depends on the 
sign of phase $\phi$. If the CSL is induced by the external field 
through Eq.~\eqref{hamChi}, the sign of phase $\phi$ should depend 
on the direction of field, thus the sign of thermal Hall conductivity 
also depends on the fields. In fact, a CSL may also be realized without 
applied fields and the quantized thermal Hall effect can be an 
intrinsic character. Overall, the quantized thermal Hall conductivity 
would be a particularly sharp signature for CSL. It is worth to note 
that the half-integer quantized therma Hall effect has been 
reported~\cite{Kasahara2018} in the Kitaev QSL candidate 
$\alpha$-RuCl$_3$ and is proposed to correspond to the 
chiral Majorana fermion edge mode.

\section{Discussion}
\label{sec5}

In summary, we have highlighted three experimental schemes 
that would give smoking-gun signatures of QSLs including 
the thermal transport and spectrum measurements, 
and successfully applied them to square lattice QSLs. 
The $\pi$-flux QSL states would result in an enhanced 
spectral periodicity of the spinon continuum. It is the 
translation symmetry that is intrinsically fractionalized 
and renders such an enhanced spectral periodicity, much 
analogous to the fractional charge excitation in the 
fractional quantum Hall states where the global U(1) 
charge conservation gives the fractional charge quantum 
number to the fractionalized excitation. Under moderate 
magnetic fields when the description of QSL is still valid, 
the Zeeman coupling will enter the spinon Hamiltonian and 
lead to an X-shaped crossing of spectrum around $\Gamma$ point, 
which is well compatible with the splitting bands for spin-$\uparrow$, 
and spin-$\downarrow$ spinons and is hard to be mimiked by 
the spin glass or disorder-induced states. Finally, if a 
CSL is realized, it would exhibit a quantized thermal Hall 
effect and a gapped spectrum. All of these sharp signatures 
can be robust even when the gauge fluctuations are included 
and do not depend on the geometry of underlying lattice, 
therefore we conclude these signatures could generally 
apply to other relevant QSLs.

\section*{Acknowledgments}

This work is supported by research 
funds from the Ministry of Science and Technology of China with 
grant No.2016YFA0301001, No.2018YFGH000095 and No.2016YFA0300500,
from Shanghai Municipal Science and Technology Major Project with 
grant No.2019SHZDZX04, and from the Research Grants Council of 
Hong Kong with General Research Fund Grant No.17303819.

\bibliography{Ref.bib}

\end{document}